\documentclass[twocolumn,showpacs,preprintnumbers,amsmath,amssymb,superscriptaddress]{revtex4}

\input epsf

\def\bea{\begin{eqnarray}}
\def\eea{\end{eqnarray}}
\def\ben{\begin{equation}}
\def\een{\end{equation}}
\def\benu{\begin{enumerate}}
\def\enu{\end{enumerate}}

\def\n{\rho}

\def\sss{\scriptscriptstyle\rm}

\def\1var{(\bx_1...\bx\N)}

\def\half{\frac{1}{2}}

\def\br{{\bf r}}

\def\bx{{x}}

\def\h{{\sss HOMO}}

\def\x{_{\sss X}}

\def\xc{_{\sss XC}}
\def\Hx{_{\sss HX}}
\def\Hxc{_{\sss HXC}}

\def\N{_{\sss N}}
\def\H{_{\sss H}}

\def\sph_int{ {\int d^3 r}}

\begin{document}


\title{{Continuum states from time-dependent density functional theory}}

\author{Adam Wasserman}
\affiliation{Department of Chemistry and Chemical Biology, Rutgers University, 610 Taylor Rd., Piscataway, NJ 08854, USA}
\author{Neepa T. Maitra}
\affiliation{Department of Physics and Astronomy,
Hunter College of the City University of New York, 695 Park Av., New York, NY 10021, USA}
\author{Kieron Burke}
\affiliation{Department of Chemistry and Chemical Biology, Rutgers University, 610 Taylor Rd., Piscataway, NJ 08854, USA}

\begin{abstract}

Linear response time-dependent density functional theory is used to
study low-lying electronic continuum states of targets that can bind
an extra electron. Exact formulas to extract scattering amplitudes
from the susceptibility are derived in one dimension. A single-pole
approximation for scattering phase shifts in three dimensions is shown
to be more accurate than static exchange for singlet electron-He$^+$
scattering.

\end{abstract}

\maketitle

\begin{section}{Introduction}

Ground-state density functional theory (DFT) \cite{HK64,KS65} has
become a popular electronic structure method in both quantum chemistry
and solid-state physics, because modern approximations produce useful
accuracy at moderate computational cost \cite{DG90,KHb00}.  Now,
electronic excitation energies of atoms and molecules are being
calculated using linear response time-dependent density functional
theory (TDDFT)\cite{RG84,PGG96}.
In this scheme, bound $\to$ bound
transition energies are first approximated by the poles of the
frequency-dependent Kohn-Sham (KS) density 
response function\cite{PGG96,C96,AGB03}, and
then corrected to the poles of the true response function,
i.e., the true excitations. Bound $\to$ continuum transitions,
however, have not been treated in the same way because branch cuts of
the KS and interacting response functions overlap, and because it is the phase
shifts, rather than the energies, that are of interest in the
scattering regime.

Even though photoresponse was addressed in the early days of TDDFT
\cite{ZS80}, and there has been a long interest of using density
functional methods for the scattering problem (e.g.\cite{TG05}), there
is no formal theory based on TDDFT to study electron scattering.  Such
a theory might be particularly useful in the emergent field of
electron-impact chemistry \cite{HGDS03}, in which large targets are
struck by low energy electrons, so that bound-free correlations are
significant \cite{N00}.

Several results relevant to this goal are presented here.
First, we provide a proof of principle: the time-dependent
response of an $N$-electron ground state contains the scattering
information for an electron scattering from the $(N-1)$-electron
target, and this is accessible via TDDFT. Second, we show how this
leads to practical ways of calculating scattering phase shifts or, in
one dimension, transmission amplitudes. 
Finally, in the simplest case, singlet scattering from He$^+$, we find that TDDFT yields
better results than static exchange, demonstrating its higher accuracy
at low computational cost.

Although we are not presenting a {\em complete} theory of electron
scattering within TDDFT, such a theory can be built upon the
rigorous results presented here, and become a competitive alternative
to existing techniques for calculating electron-molecule scattering
cross sections (e.g. \cite{WM96}). Since continuum states are the current-carrying states in molecular electronic devices, we also anticipate applications of our 1-d results in the field of electronic transport through molecular wires \cite{VL02}.

\end{section}

\begin{section}{Extracting scattering information from the susceptibility}
\begin{subsection}{Theory}

Our starting point is the Dyson-like response equation that
relates the susceptibility $\chi(\br,\br';\omega)$ of a system of $N$
interacting electrons with that of its ground-state KS analog,
$\chi_s(\br,\br';\omega)$ \cite{PGG96}. In operator form ($*$
indicates spatial and spin convolution): \ben
\chi=\chi_s+\chi_s*f\Hxc{*}\chi~~, \label{Dyson_time} \een where
$f\Hxc$ is the Hartree-exchange-correlation kernel (we use atomic
units throughout): \ben
f\Hxc[\n](\br,\br';t-t')\equiv\frac{\delta(t-t')}{|\br-\br'|}+\left.\frac{\delta{v\xc(\br,t)}}{\delta{\n(\br',t')}}\right|_{\n}~~,
\label{fxc_time} \een a functional of the $N$-electron
ground-state density $\n(\br)$. In Eq.(\ref{fxc_time}), $v\xc(\br,t)$
is the time-dependent exchange-correlation potential induced when
a time-dependent perturbation is applied to the $N$-electron
ground state.
We write the spin-decomposed susceptibility in the Lehman
representation: \ben
\chi_{\sigma\sigma'}(\br,\br';\omega)=\left[\sum_n\frac{F_{n\sigma}(\br)F_{n\sigma'}^*(\br')}{\omega-\Omega_n+i0^+}+cc(\omega\rightarrow-\omega)\right]~,
\label{Lehman} \een with \ben
F_{n\sigma}(\br)=\langle\Psi_0|\hat{\n}_{\sigma}(\br)|\Psi_n\rangle~~;~~\hat{\n}_{\sigma}(\br)=\sum_{i=1}^{N}\delta(\br-\hat{\br}_i)\delta_{\sigma\hat{\sigma}_i}
\label{f} \een where $\Psi_0$ is the ground state of the
$N$-electron system, $\Psi_n$ its $n^{\rm{th}}$ excited state, and
$\hat{\n}_{\sigma}(\br)$ is the $\sigma$-spin density operator. In
Eq.(\ref{Lehman}), $\Omega_n$ is the $\Psi_0\rightarrow\Psi_n$
transition frequency.

For the remainder of this section, we restrict the analysis to one dimension.
Consider large distances, where the $N$-electron
ground-state density is dominated by the decay of the highest
occupied KS orbital \cite{KD80}; the ground-state wavefunction
behaves as \cite{EBP96}: \ben
\Psi_0\mathop{\rightarrow}_{x\rightarrow\infty}\psi_0^{N-1}(x_2,...x_{N})\sqrt{\frac{\n(x)}{N}}S_0(\sigma,\sigma_2,...\sigma_{N})\\
\label{psi_0} \een where $\psi_0^{N-1}$ is the ground-state
wavefunction of the $(N-1)$-electron system (the {\em target}),
$S_0$ the spin function of the ground state and $\n(x)$ the
$N$-electron ground-state density. Similarly, \ben
\Psi_n\mathop{\rightarrow}_{x\rightarrow\infty}\psi_{n_t}^{N-1}(x_2,...x_{N})\frac{\phi_{k_n}(x)}{\sqrt{N}}S_n(\sigma,\sigma_2,...\sigma_{N})
\label{psi_n} \een where $\psi_{n_t}^{N-1}$ is an eigenstate of the
target (labeled by $n_t$), $S_n$ is the spin function of the
$n^{\rm{th}}$ excited state, and $\phi_{k_n}(x)$ a one-electron
orbital.

We focus on elastic scattering, so the contribution
to $F_{n\sigma}(x)$ from channels where the target is excited vanishes as $x\rightarrow\infty$ due to orthogonality.
Inserting
Eqs.(\ref{psi_0}) and (\ref{psi_n}) into the 1d-version of Eq.(\ref{f}), and taking into
account the antisymmetry of both $\Psi_0$ and $\Psi_n$,
\begin{eqnarray}
\nonumber
\lefteqn{
F_{n\sigma}(x)\mathop{\rightarrow}_{x\rightarrow\infty}\sqrt{\n(x)}\phi_{k_n}(x)\delta_{0,n_t}}
\\
&&\times\sum_{\sigma_2...\sigma_{N}}{S_0^*(\sigma...\sigma_{N})S_n(\sigma...\sigma_{N})}
\label{f_largex}
\end{eqnarray}
The susceptibility at large distances is then obtained by inserting Eq.(\ref{f_largex}) into Eq.(\ref{Lehman}):
\begin{eqnarray}
\nonumber
\lefteqn{
\chi(x,x';\omega)=\sum_{\sigma\sigma'}\chi_{\sigma\sigma'}(x,x';\omega)\mathop{\rightarrow}_{x,x'\rightarrow\pm\infty}\sqrt{\n(x)\n(x')}}
\\
&&\times\sum_n\frac{\phi_{k_n}(x)\phi_{k_n}^*(x')}{\omega-\Omega_n+i\eta}\delta_{0,n_t}\delta_{S_0,S_n}+cc(\omega\rightarrow-\omega)
\label{chi_largex}
\end{eqnarray}
Since only scattering states of the $N$-electron optical potential
contribute to the sum in Eq.(\ref{chi_largex}) at large distances,
it becomes an integral over wavenumbers $k=\sqrt{2\varepsilon}$,
where $\varepsilon$ is the energy of the projectile electron: \ben
\sum_n\frac{\phi_{k_n}(x)\phi_{k_n}^*(x')}{\omega-\Omega_n+i\eta}\mathop{\rightarrow}_{x,x'\rightarrow\pm\infty}\frac{1}{2\pi}\int_{0\rm[R],[L]}^{\infty}{\frac{\phi_k(x)\phi_k^*(x')}{\omega-\Omega_k+i\eta}dk}
\label{integral} \een In this notation, the functions $\phi_{k_n}$
are box-normalized, and $\phi_{k_n}(x)=\phi_k(x)/\sqrt{L}$, where
$L\rightarrow\infty$ is the length of the box. The transition
frequency $\Omega_n=E_n^{N}-E_0^{N}$ is now simply
$\Omega_k=E_0^{N-1}+k^2/2-E_0^{N}=k^2/2+I$, where $I$ is the first
ionization potential of the $N$-electron system, and $E_0^M$ and
$E_n^M$ are the ground and $n^{th}$ excited state energies of the
$M$-electron system. The subscript ``[R],[L]'' implies that the
integral is over both orbitals satisfying {\em right} and {\em
left} boundary conditions: \ben
\phi_k^{\stackrel{\mbox{\tiny{[R]}}}{\mbox{\tiny{[L]}}}}(x)\rightarrow\left\{\begin{array}{l
l}e^{\pm{ikx}}+r_ke^{\mp{ikx}}&,~~x\rightarrow\mp\infty\\t_ke^{\pm{ikx}}&,~~x\rightarrow\pm\infty\end{array}\right.
\label{R_and_L} \een

When $x\rightarrow-\infty$ and $x'=-x$ the integral of
Eq.(\ref{integral}) is dominated by a term that oscillates in
space with wavenumber $2\sqrt{2(\varepsilon-I)}$ and amplitude
given by the transmission amplitude for spin-conserving collisions
$t_k$ at that wavenumber. Denoting this by $\chi^{\rm osc}$, we obtain:

\ben
t(\varepsilon)=\lim_{x\rightarrow-\infty}\left[\frac{i\sqrt{2\varepsilon}}{\sqrt{\n(x)\n(-x)}}\chi^{\rm
osc}(x,-x;\varepsilon+I)\right]~~. \label{t_from_chi} \een While this
formula also applies to the KS system, its transmission
$t_s(\varepsilon)$ can be easily obtained by solving a {\em potential
scattering} problem (i.e., scattering off the $N$-electron ground-state
KS potential). The exact amplitudes $t(\varepsilon)$ of the many-body
problem are formally related to the $t_s(\varepsilon)$ through
Eqs.(\ref{t_from_chi}) and (\ref{Dyson_time}). This is the main result of this work: the time-dependent response of the $N$-electron
ground-state contains the scattering information, and is
accessible via TDDFT. A potential scattering problem is solved first
for the $N$-electron ground-state KS potential, and the scattering
amplitudes thus obtained are further corrected by $f\Hxc$ to account
for, e.g., polarization effects.

While
Eq.(\ref{t_from_chi}) seems impractical as a basis for
computations, it leads to practical approximations. For example, if Eq.(\ref{Dyson_time}) is iterated once, we find through Eq.(\ref{t_from_chi}) the
following useful distorted-wave-Born-type approximation for the transmission amplitude: 
\ben t(\varepsilon)=t_s(\varepsilon)+\frac{1}{i\sqrt{2\varepsilon}}\langle\langle {\sss{\rm
HOMO}},\varepsilon|\hat{f}\Hxc(\varepsilon+I)|{\sss {\rm
HOMO}},\varepsilon\rangle\rangle~~,\label{BA_for_t}\een where $|{\sss{\rm
HOMO}},\varepsilon\rangle\rangle$ is the product of the highest occupied KS
orbital and the continuum KS orbital of energy $\varepsilon$.
\end{subsection}
\begin{subsection}{Example}

We illustrate on a simple 1-d model of an electron scattering from a one-electron atom of nuclear charge $Z$ \cite{R71} in the weak interaction limit:
\ben
\hat{H}=-\frac{1}{2}\frac{d^2}{dx_1^2}-\frac{1}{2}\frac{d^2}{dx_2^2}-Z\delta(x_1)-Z\delta(x_2)+\lambda\delta(x_1-x_2)~~,
\label{H}
\een
Electrons interact via a delta-function repulsion, scaled by $\lambda$. With $\lambda=0$ the ground state density is a simple exponential, analogous to hydrogenic atoms in 3d.

(i) {\em Exact solution in the weak interaction limit:} First, we solve for the exact transmission amplitudes to
first order in $\lambda$ using the static exchange method
~\cite{BJ83}.
The results for triplet ($t_{trip}$) and singlet ($t_{sing}$)
scattering are: \begin{eqnarray} \nonumber
t_{trip}={t_0}~~&,&~~t_0=\frac{ik}{Z+ik}~~~~\\
\label{t_exact}
t_{sing}=t_{0}+2\lambda{t_1}~~&,&~~t_1=\frac{-ik^2}{(k-iZ)^2(k+iZ)}
\end{eqnarray}

 (ii) {\em Our TDDFT solution:} The ground-state of the $N$-electron system ($N=2$) is given to ${\cal O}(\lambda)$ by: \ben
\Psi_0(x_1\sigma_1,x_2\sigma_2)=\frac{1}{\sqrt{2}}\phi_0(x_1)\phi_0(x_2)\left[\delta_{\sigma_1\uparrow}\delta_{\sigma_2\downarrow}-\delta_{\sigma_1\downarrow}\delta_{\sigma_2\uparrow}\right]~~,
\label{gs} \een where the orbital $\phi_0(x)$ satisfies
\cite{LSY03,Rudy}: \ben
\left[-\frac{1}{2}\frac{d^2}{dx^2}-Z\delta(x)+\lambda|\phi_0(x)|^2\right]\phi_0(x)=\mu\phi_0(x)
\label{HF_eqn} \een To first order in $\lambda$, \ben
\phi_0(x)=\sqrt{Z}e^{-Z|x|}+\frac{\lambda}{8\sqrt{Z}}\left(2e^{-3Z|x|}+e^{-Z|x|}(4Z|x|-3)\right)
\label{phi0} \een
The bare KS transmission amplitudes $t_s(\varepsilon)$
characterize the asymptotic behavior of the continuum states of
$v_s(x)=-Z\delta(x)+\lambda|\phi_0(x)|^2~~$, and can be obtained
to ${\cal O}(\lambda)$ by a distorted-wave Born approx.
(see e.g. Ref.\cite{Friedrich}): \ben t_s=t_0+\lambda{t_1}
\label{ts} \een The result is plotted in Fig.1, along
with the interacting singlet and triplet transmission amplitudes, Eqs.(\ref{t_exact}).

We now apply Eq.(\ref{t_from_chi}) to show that the $f\Hxc$-term of Eq.(\ref{Dyson_time}) corrects the $t_s$ values to their exact singlet and triplet amplitudes.
\begin{figure}
\epsfxsize=70mm \epsfbox{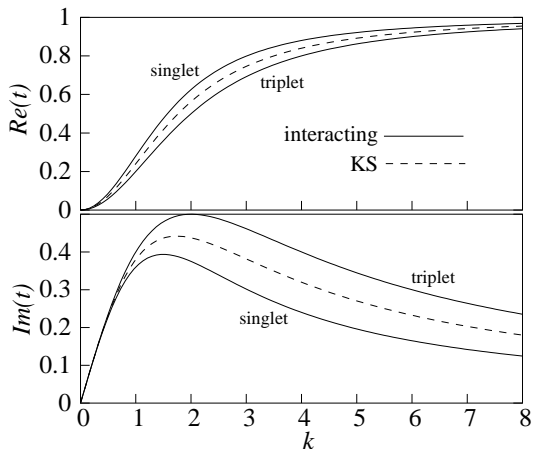}
\caption{Real and imaginary parts of the KS transmission amplitude
$t_s$, and of the interacting singlet and triplet amplitudes (to first order in $\lambda$), for
the model system of Eq.(\ref{H}). $Z=2$ and $\lambda$=0.5 in this
plot.} \label{f:t}
\end{figure}

We need $f\Hxc$ only to ${\cal O}(\lambda)$: \ben
f\Hx^{\sigma\sigma'}(x,x';\omega)=\lambda\delta(x-x')(1-\delta_{\sigma\sigma'})~~,
\label{fhxc} \een where the $f\Hxc$ of Eq.(\ref{Dyson_time}) is
given to ${\cal O}(\lambda)$ by
$f\Hx=f\H+f\x=\frac{1}{4}\sum_{\sigma\sigma'}{f\Hx^{\sigma\sigma'}}$ ($=\half{f\H}$ here). Eq.(\ref{fhxc}) yields: \ben
\chi(x,x';\omega)=\chi_s(x,x';\omega)+\frac{\lambda}{2}\int{dx''\chi_s(x,x'';\omega)\chi(x'',x';\omega)}
\label{dyson_chi} \een Since the ground state of the $N$-electron
system is a spin-singlet, the Kronecker delta $\delta_{S_0,S_n}$
in Eq.(\ref{chi_largex}) implies that only $singlet$ scattering
information may be extracted from $\chi$, whereas information
about triplet scattering requires the magnetic susceptibility
${\cal{M}}=\sum_{\sigma\sigma'}(\sigma\sigma')\chi_{\sigma\sigma'}$,
related to the KS susceptibility by spin-TDDFT~\cite{PG96}: \ben
{\cal{M}}(x,x';\omega)=\chi_s(x,x';\omega)-\frac{\lambda}{2}\int{dx''\chi_s(x,x'';\omega){\cal{M}}(x'',x';\omega)}
\label{dyson_chimag} \een

For either singlet or triplet case, since the correction to
$\chi_s$ is multiplied by $\lambda$, the leading
correction to $t_s(\varepsilon)$ is determined by the same
quantity, $\hat{\chi}_s^{(0)}*\hat{\chi}_s^{(0)}$, where
$\hat{\chi}_s^{(0)}$ is the $0^{\rm th}$ order approximation to
the KS susceptibility (i.e. with
$v_s(x)=v_s^{(0)}(x)=-Z\delta(x)$). Its oscillatory part at large
distances \cite{MWB03} (multiplied by $\sqrt{\n(x)\n(-x)}/ik$, see
Eq.(\ref{t_from_chi})) is equal to $\lambda{t_1}$. We then find
through
Eqs.(\ref{t_from_chi}), (\ref{dyson_chi}), and
(\ref{dyson_chimag}) that \ben
t_{sing}=t_s+\lambda{t_1}~~,~~t_{trip}=t_s-\lambda{t_1}~~,
\label{tsinglet} \een in agreement with
Eqs.(\ref{t_exact}).

\end{subsection}
\end{section}

\begin{section}{Single pole approximation  in the continuum}

We have yet to prove an analog of Eq.(\ref{t_from_chi}) for Coulomb
repulsion in three dimensions. But here we use quantum-defect theory
\cite{S58} to deduce the result at zero energy. Consider the $l=0$
Rydberg series of bound states converging to the first ionization
threshold $I$ of the $N$-electron system: \ben
E_n-E_0=I-1/\left[2(n-\mu_n)^2\right]~~, \een where $\mu_n$ is the
quantum defect of the $n^{th}$ excited state. Let \ben
\epsilon_{n}=-1/\left[2(n-\mu_{s,n})^{2}\right]~~, \een be the KS
orbital energies of that series. The true transition
frequencies $\omega_n=E_n-E_0$, are related through TDDFT to the KS
frequencies $\omega_{s,n}=\epsilon_n-\epsilon_{\h}$. Within the
single-pole approximation (SPA) \cite{PGG96}: \ben
\omega_n=\omega_{s,n}+2\langle\langle{\h{,n}}|\hat{f}\Hxc(\omega_n)|{\h{,n}}\rangle\rangle
\label{SPA} \een Numerical studies \cite{ARU98} suggest that
$\Delta\mu_n=\mu_n-\mu_{s,n}$ is a small number when $n\to\infty$.
Expanding $\omega_n$ around $\Delta\mu_n=0$, and using
$I=-\epsilon_{\h}$, we find: \ben
\omega_n=\omega_{s,n}-\Delta\mu_n/(n-\mu_{s,n})^3
\label{compareSPA} \een We conclude that, within the SPA, \ben
\Delta\mu_n=-2(n-\mu_{s,n})^3\langle\langle{\h,n}|\hat{f}\Hxc(\omega_n)|{\h,n}\rangle\rangle.
\label{delta_mu} \een Letting $n\to\infty$, Seaton's theorem
($\pi\lim_{n\to\infty}\mu_{n}=\delta(\varepsilon\to 0^+)$)\cite{S58}
implies: \ben
\delta(\varepsilon)=\delta_s(\varepsilon)-2\pi\langle\langle{\h,\varepsilon}|\hat{f}\Hxc(\varepsilon+I)|{\h,\varepsilon}\rangle\rangle
\label{SPA_for_eta} \een a relation for the phase-shifts $\delta$ in
terms of the KS phase-shifts $\delta_s$ applicable when $\varepsilon\to 0^+$. The factor
$(n-\mu_{s,n})^3$ of Eq.({\ref{delta_mu}}) gets absorbed into the energy-normalization factor of the KS continuum states. 

\begin{figure}
\epsfxsize=70mm \epsfbox{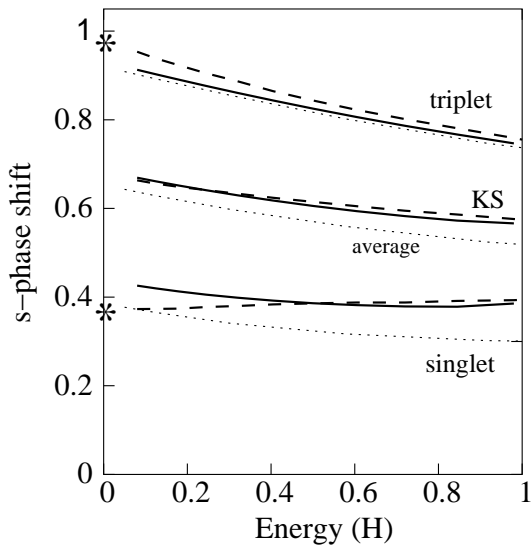}
\caption{$s$-phase shifts as a function of energy for electron scattering from He$^+$. {\em Dashed lines}: the line labeled KS corresponds to the phase shifts from the {\em exact} KS potential of the He atom; the other dashed lines correspond to the TDDFT singlet and triplet phase shifts calculated in the present work according to Eq.(\ref{SPA_for_eta}). {\em Solid lines}: accurate wavefunction calculations of electron-He$^+$ scattering from Ref.\cite{B02}. The solid line in the center is the average of singlet and triplet phase shifts. {\em Dotted lines}: Static exchange calculations, from Ref.\cite{LM80}. The asterisks at zero energy correspond to extrapolating the bound $\to$ bound results of Ref.\cite{BPG00}.}
\label{f:real}
\end{figure}

We illustrate in Fig.2 the remarkable accuracy of
Eq.(\ref{SPA_for_eta}) when applied to the case of electron scattering
from He$^+$. For this system, an essentially exact ground-state
potential for the $N=2$ electron system is known. This was found by
inverting the KS equation using the ground-state density of an
extremely accurate wavefunction calculation of the He
atom~\cite{UG94}.  We calculated the low-energy KS $s$-phase shifts
from this potential, $\delta_s(\varepsilon)$ (dashed line in the
center, Fig.2), and then corrected these phase shifts
according to Eq.(\ref{SPA_for_eta}) employing the BPG approximation to
$f\Hxc$ \cite{BPG00} (= adiabatic local density approximation for
antiparallel contribution to $f\Hxc$ and exchange-only approximation
for the parallel contribution). We also plot the results of a recent
highly accurate wavefunction calculation \cite{B02} (solid), and of
static-exchange calculations \cite{LM80} (dotted).  The results show
that phase shifts from the $N$-electron ground-state KS potential,
$\delta_s(\varepsilon)$, are an excellent approximation to the average
of the true singlet/triplet phase shifts for an electron scattering
from the $(N-1)$-electron target, just as in our one-dimensional
model; they also show that TDDFT, with existing approximations, works
very well to correct scattering from the KS potential to the true
scattering phase shifts, at least at low energies. In fact, for the
singlet phase shifts, TDDFT does better than the computationally more
demanding static exchange method, and for the triplet case TDDFT does
only slightly worse. Even though Eq.(\ref{SPA_for_eta}) is, strictly
speaking, only applicable at zero energy (marked with asterisks in
Fig.2), it clearly provides a good description for finite
(low) energies. It is remarkable that the antiparallel spin kernel,
which is completely local in space and time, and whose value at each
point is given by the exchange-correlation energy density of a uniform
electron gas (evaluated at the ground-state density at that point),
yields phase shifts for e-He$^+$ scattering with less than 20\%
error. Since a signature of density-functional methods is that, with
the same functional approximations, exchange-correlation effects are
often better accounted for in larger systems, the present approach
holds promise as a practical method for studying large targets.

\end{section}

\begin{section}{Conclusion}

To summarize, we have shown how, in one-dimension, scattering
amplitudes may be obtained from TDDFT, and deduced the results for
three-dimensions near zero energy for Coulombic systems. The ultimate
goal is to accurately treat bound-free correlation for low energy
electron scattering from polyatomic molecules, with a computational
cost lower than that of static exchange.  An obvious limitation of the
present approach is that it can only be applied to targets than
bind an extra electron, and there is much work yet to be done: general
proof of principle in three dimensions, testing of the accuracy of
approximate ground-state KS potentials, developing and testing
approximate solutions to the TDDFT Dyson-like equation, extending the
methodology to cases where the anion has a sharp resonance rather than
a ground state, etc.

\end{section}

\begin{section}*{Acknowledgements}

We thank Michael Morrison for inspiring discussions. This work was
supported by the Petroleum Research Fund grant No. 36515-AC6, DOE grant No. DE-FG02-01ER45928, and NSF grant No.CHE-0355405. 

\end{section}

\pagebreak

%
%
%
%
%
%








\end{document}